\documentclass[showpacs,preprintnumbers,amsmath,amssymb]{revtex4}
\usepackage{amsmath}
\usepackage{graphicx}
\usepackage{subfig}
\usepackage{hyperref}
\usepackage{amssymb}
\usepackage[english]{babel}
\usepackage{epsfig}
\usepackage{wasysym}

\begin{document}

\title{ Generalized Teleparallel Theory\\}

\author{ Ednaldo L. B. Junior$^{(a,b)}$\footnote{E-mail address:ednaldobarrosjr@gmail.com}, Manuel E. Rodrigues$^{(a,c)}$\footnote{E-mail address: esialg@gmail.com}  }
\affiliation{$^{a}$ \, Faculdade de F\'{\i}sica, PPGF, Universidade Federal do Par\'{a}, 66075-110, Bel\'{e}m, Par\'{a}, Brazil.\\
$^b$\ Faculdade de Engenharia da Computa\c{c}\~{a}o, Universidade Federal do Par\'{a}, Campus Universitário de Tucuru\'{\i}, CEP: 68464-000, Tucuru\'{\i}, Par\'{a}, Brazil.\\
$^c$\ Faculdade de Ci\^{e}ncias Exatas e Tecnologia, Universidade Federal do Par\'{a}\\
Campus Universit\'{a}rio de Abaetetuba, CEP 68440-000, Abaetetuba, Par\'{a}, Brazil.}


\begin{abstract}
We construct a theory in which the gravitational interaction is described only by torsion, but that generalizes the Teleparallel Theory still keeping the invariance of local Lorentz transformations in one particular case. We show that our theory falls, to a certain limit of a real parameter, in the $f(\bar{R})$ Gravity or, to another limit of the same real parameter, in a modified $f(T)$ Gravity, interpolating between these two theories and still can fall on several other theories. We explicitly show the equivalence with $f(\bar{R})$ Gravity for cases of Friedmann-Lemaitre-Robertson-Walker flat metric for diagonal tetrads, and a metric with spherical symmetry for diagonal and non-diagonal tetrads. We do still four applications, one in the reconstruction of the de Sitter universe cosmological model, for obtaining a static spherically symmetric solution type-de Sitter for a perfect fluid,   for evolution of the state parameter $\omega_{DE}$ and for the thermodynamics to the apparent horizon.
\end{abstract}

\date{\today}

\maketitle



\section{Introduction}
\label{sec1}
One of the most important events in modern physics is that our universe is expanding accelerated \cite{Ia}. However, a plausible explanation for this is commonly done using the model of a very exotic fluid called dark energy, which has negative
pressure. Another well-known possibility is to modify Einstein's General Relativity (GR) \cite{plebanski}, making the action of the theory depend on a function of the curvature scalar $R$, but at a certain limit of parameters the theory falls on GR. This way to explain the accelerated expansion of our universe is known as Modified Gravity or Generalized. Considering that the gravitational interaction is described only by the curvature of space-time, we can generalize the Einstein-Hilbert action through analytic function of scalars of the theory, as for example the gravities $f(\bar{R})$ \cite{fR}, with $\bar{R}$ being the Ricci scalar or curvature scalar,  $f(\bar{R},\Theta)$ \cite{fRT}, with $\Theta$ being the trace of energy-momentum tensor, or yet $f(G)$ \cite{fG}, $f(\bar{R},G)$ \cite{fRG} and $f(\bar{R},\Theta,\bar{R}_{\mu\nu}\Theta^{\mu\nu})$ \cite{odintsov}, with $\Theta^{\mu\nu}$ being the energy-momentum tensor.
\par 
An alternative to consistently describe the gravitational interaction is one which only considers the torsion of space-time, thus cancelling out any effect of the curvature. This approach is known as Teleparallel Theory (TT) \cite{TT}, which is demonstrably equivalent to GR. In order to describe not only the gravitational interaction, but also the accelerated expansion of our universe, Ferraro and Fiorini \cite{ferraro}  proposed a possible generalization of the TT, which became known as $f(T)$ Gravity \cite{fT}, in which up to now has provided good results in both cosmology as local phenomena of gravitation. A key problem in $f(T)$ Gravity is that it breaks the invariance under local Lorentz transformations complicating the interpretation of the relationship between all inertial frames of the tangent space to the differentiable manifold (space-time) \cite{sotiriou}. This problem may lead to the emergence of new degrees of freedom spurious who are responsible for the breakdown of the local Lorentz symmetry \cite{li}. A consequence of the formulated theory using a scalar which is not invariant by local Lorentz transformations, the torsion scalar $T$ in this case, is that instead of the theory presenting differential equations of motion of fourth order, as in the case of the $f(\bar{R})$ Gravity, it has second-order differential equations. That seems like a benefit but is a consequence of this fact on the local Lorentz symmetry. We still have which  this generalization of the TT is not equivalent to generalization $f(\bar{R})$ for RG.
\par 
This is the main reason that will address the construction of a theory that generalize the TT, but which still keep the local Lorentz symmetry  on a particular case. Therefore, it is clear that we must build the function of action with dependence on a scalar that at some limit is invariant under local Lorentz transformations. It will be shown soon forward.
\par 
The paper is organized as follows. In section \ref{sec2} we do a review of $f(T)$ Gravity, introducing the functional variation method used in this work, obtaining the equations of motion of this theory, noting a poorly treated point at the limit to GR. In section \ref{sec3} we propose the action of Generalized Teleparallel Theory, we obtain the equations of motion through functional variation of the same and compared with $f(T)$ Gravity. We show the equivalence of our theory with $f(\bar{R})$ Gravity, in the case of cosmology for the line  element of flat FLRW metric in subsection \ref{subsec4.1}, and also in the case of a spherically symmetric line element in subsection \ref{subsec4.2}. We show still the equivalence of our theory with a particular case of $f(T,B)$ Gravity in section \ref{sec5}. In section \ref{sec6} we make four applications, one where we reconstructed the action of our theory for the universe of the model of de Sitter, another where we obtain a static type-de Sitter solution;  we analyse teh evolution for the state parameter to dark energy and the thermodynamics for a cosmological model. We make our final considerations in section \ref{sec7}.

\section{The equations of motion for $f(T)$ Gravity}
\label{sec2}
The geometry of a space-time can be characterized by the curvature and torsion. In the particular case in which we only consider the curvature and torsion being zero, we have defined, together with the metricity condition $\nabla_{\mu}g_{\alpha\beta}\equiv 0$ where $g_{\alpha\beta}$ are the components of the metric tensor, a Riemannian geometry where the connection $\bar{\Gamma}^{\mu}_{\;\;\alpha\beta}$ is symmetric in the last two indices. Already in the particular case that we consider only torsion (Riemann tensor identically zero, case without curvature) in the space-time, we can then work with objects that depend solely on the so-called tetrads matrices and its derivatives as dynamic fields. 
\par 
In the space-time having only torsion, the line element can be represented through two standard forms
\begin{eqnarray}
dS^2=g_{\mu\nu}dx^{\mu}dx^{\nu}=\eta_{ab}\theta^{a}\theta^{b}\,,\label{le}
\end{eqnarray}
where we have the following relationships $g_{\mu\nu}=\eta_{ab}e^{a}_{\;\;\mu}e^{b}_{\;\;\nu}$, $g^{\mu\nu}=\eta^{ab}e_{a}^{\;\;\mu}e_{b}^{\;\;\nu}$, $\theta^{a}=e^{a}_{\;\;\mu}dx^{\mu}$, $e^{a}_{\;\;\mu}e_{a}^{\;\;\nu}=\delta^{\nu}_{\mu}$ e $e^{a}_{\;\;\mu}e_{b}^{\;\;\mu}=\delta^{a}_{b}$, with $e^{a}_{\;\;\mu}$ being the tetrads matrices and $e_{a}^{\;\;\mu}$ its inverse, and $[\eta_{ab}]=diag[1,-1,-1,-1]$ the Minkowski metric. We adopt the Latin indices for the tangent space and the Greeks into space-time.
\par 
We will first establish the equations of motion for the theory $f(T)$, thus showing that the functional variation method adopted here is consistent.
\par
We restrict the geometry to of Weitzenbock where we have the following connection
\begin{eqnarray}
\Gamma^{\sigma}_{\;\;\mu\nu}=e_{a}^{\;\;\sigma}\partial_{\nu}e^{a}_{\;\;\mu}=-e^{a}_{\;\;\mu}\partial_{\nu}e_{a}^{\;\;\sigma}\label{WC}\; .
\end{eqnarray}
All Riemann tensor components are identically zero for the connection (\ref{WC}). We can then define the components of the tensor of torsion and contortion as
\begin{eqnarray}
T^{\sigma}_{\;\;\mu\nu}&=&\Gamma^{\sigma}_{\;\;\nu\mu}-\Gamma^{\sigma}_{\;\;\mu\nu}=e_{a}^{\;\;\sigma}\left(\partial_{\mu} e^{a}_{\;\;\nu}-\partial_{\nu} e^{a}_{\;\;\mu}\right)\label{tor}\;,\\
K^{\mu\nu}_{\;\;\;\;\alpha}&=&-\frac{1}{2}\left(T^{\mu\nu}_{\;\;\;\;\alpha}-T^{\nu\mu}_{\;\;\;\;\alpha}-T_{\alpha}^{\;\;\mu\nu}\right)\label{cont}\; .
\end{eqnarray} 
We can also define a new tensor, so we write a more elegant way the equations of motion, through the components of the tensor torsion and contortion, as
\begin{eqnarray}
S_{\alpha}^{\;\;\mu\nu}=\frac{1}{2}\left( K_{\;\;\;\;\alpha}^{\mu\nu}+\delta^{\mu}_{\alpha}T^{\beta\nu}_{\;\;\;\;\beta}-\delta^{\nu}_{\alpha}T^{\beta\mu}_{\;\;\;\;\beta}\right)\label{s}\;.
\end{eqnarray}
We define the torsion scalar as
\begin{eqnarray}
T=T^{\alpha}_{\;\;\mu\nu}S_{\alpha}^{\;\;\mu\nu}=\frac{1}{4}T^{\alpha}_{\;\;\mu\nu}T_{\alpha}^{\;\;\mu\nu}+\frac{1}{2}T^{\alpha}_{\;\;\mu\nu}T^{\nu\mu}_{\;\;\;\;\alpha}-T^{\alpha}_{\;\;\mu\alpha}T^{\beta\mu}_{\;\;\;\;\beta}\label{ts}\,.
\end{eqnarray}
Some observations are important here. The first is that there is a direct analogy to a space only with torsion and another considering only curvature in that the connections are related by
\begin{eqnarray}
\bar{\Gamma}^{\alpha}_{\;\;\mu\nu}=\Gamma^{\alpha}_{\;\;\mu\nu}-g_{\mu\lambda}K^{\alpha\lambda}_{\;\;\;\;\nu}\label{conecr}\,,
\end{eqnarray}
where $\bar{\Gamma}^{\alpha}_{\;\;\mu\nu}$ is the Levi-Civita connection, which is symmetric in the last two indices. The second observation is that the torsion scalar $T$ is not a Lorentz scalar (in the tangent space), being only a scalar in the tensorial indices (space-time) \cite{aldrovandi}. This is precisely the cause for that theory built starting this scalar breaks down the invariance by local Lorentz transformations. We can in reality build the curvature scalar analog, through of the torsion scalar, to relation \cite{aldrovandi}
\begin{eqnarray}
\bar{R}=-T-2\bar{\nabla}^{\mu}T^{\alpha}_{\;\;\mu\alpha}=-T-2e^{-1}\partial_{\mu}\left(eg^{\mu\lambda}T^{\alpha}_{\;\;\lambda\alpha}\right)\label{R}\,,
\end{eqnarray}  
where $e=det[e^{a}_{\;\;\mu}]=\sqrt{-g}$, with $g=det[g_{\mu\nu}]$. The curvature scalar $\bar{R}$ in (\ref{R}) is a Lorentz scalar as well as a scalar on tensorial indices. That is why the $f (\bar{R})$ Gravity is a theory that is invariant under local Lorentz transformations and general coordinates transformations (tensorial). 
\par 
Is then possible to construct a generalization of the Teleparallel Theory (TT) using the following action of the $f(T)$ Gravity,
\begin{eqnarray}
S_{f(T)}=\int d^4x \mathcal{L}_{f(T)}=\int d^4x \left[\frac{e}{2\kappa^2}f(T)-\mathcal{L}_{matter}\right]\,\label{fTaction}
\end{eqnarray}
where $\kappa^2=8\pi G_{Newton}$, $f(T)$ is a function of the torsion scalar and $\mathcal{L}_{matter}$ is the Lagrangian density of the material content. We call attention to the true sign $(-)$ in the front of the matter term. This so far has not been explicitly addressed in the literature of this theory, because we still have few models that couple content materials that need to be obtained through functional variation in principle. This signal is essential if the theory is equivalent to GR at some limit. It will soon be clear forward.  
\par 
Making the functional variation of the action (\ref{fTaction}) we have
\begin{eqnarray}
\delta S_{f(T)}&=&\frac{1}{2\kappa^2}\int d^4x\left[f\delta e+e\delta f-2\kappa^2 \delta\mathcal{L}_{matter}\right]\,,\nonumber\\
&=&\frac{1}{2\kappa^2}\int d^4x\left[f\frac{\partial e}{\partial e^{a}_{\;\;\sigma}}\delta e^{a}_{\;\;\sigma}+e\frac{df}{dT}\delta T\right] -\int d^4x \delta\mathcal{L}_{matter}\,,\nonumber\\
&=&\delta S_T-\delta S_{matter}\label{delS1}\,,
\end{eqnarray}
with $\delta S_{matter}=\int d^4x\;\delta\mathcal{L}_{matter}$. Now let's do first the functional variation of the matter term,
\begin{eqnarray}
\delta S_{matter}&=&\int d^4x \left[\frac{\partial\mathcal{L}_{matter}}{\partial e^a_{\;\;\sigma}}\delta e^a_{\;\;\sigma}+\frac{\partial \mathcal{L}_{matter}}{\partial(\partial_\alpha e^a_{\;\;\sigma})}\delta(\partial_\alpha e^a_{\;\;\sigma})\right]\nonumber\,,
\end{eqnarray}
that making the integration by part of the latter term, considering $\delta e^{a}_{\;\;\sigma}\big|_{surface}\equiv 0$, we have
\begin{eqnarray}
\delta S_{matter}&=&\frac{1}{2\kappa^2}\int d^4x \;2\kappa^2\left[\frac{\partial\mathcal{L}_{matter}}{\partial e^a_{\;\;\sigma}}\delta e^a_{\;\;\sigma}-\partial_{\alpha}\left(\frac{\partial\mathcal{L}_{matter}}{\partial(\partial_\alpha e^a_{\;\;\sigma})}\right)\delta e^a_{\;\;\sigma}\right]\nonumber\\
&=&\frac{1}{2\kappa^2}\int d^4x\;2\kappa^2 e\Theta_a^{\;\;\sigma}\delta e^a_{\;\;\sigma}\,,\label{delSm1}
\end{eqnarray}
where $\Theta_a^{\;\;\sigma}=e_a^{\;\;\beta}\Theta_{\beta}^{\;\;\sigma}$, and we define $\Theta_\nu^{\;\;\sigma}$ as being the energy-momentum tensor.
\par 
We have now the functional variation of geometric part,
\begin{eqnarray}
\delta S_T=\frac{1}{2\kappa^2}\int d^4x\left\{f\frac{\partial e}{\partial e^{a}_{\;\;\sigma}}\delta e^{a}_{\;\;\sigma}+e\frac{df}{dT}\left[\frac{\partial T}{\partial e^{a}_{\;\;\sigma}}\delta e^{a}_{\;\;\sigma}+\frac{\partial T}{\partial (\partial_{\alpha}e^{a}_{\;\;\sigma})}\delta (\partial_{\alpha}e^{a}_{\;\;\sigma})\right]\right\} \nonumber\,.
\end{eqnarray} 
Doing integration by parts the last term, considering $\delta e^{a}_{\;\;\sigma}\big|_{surface}\equiv 0$, we obtain
\begin{eqnarray}
\delta S_T=\frac{1}{2\kappa^2}\int d^4x\left\{f\frac{\partial e}{\partial e^{a}_{\;\;\sigma}}+ef_T\frac{\partial T}{\partial e^{a}_{\;\;\sigma}}-\partial_{\alpha}\left[ef_T\frac{\partial T}{\partial (\partial_{\alpha}e^{a}_{\;\;\sigma})}\right]\right\}\delta e^{a}_{\;\;\sigma} \label{delST1}\,.
\end{eqnarray} 
where $f_T=df/dT$. Taking (\ref{delSm1}) and (\ref{delST1}) and replacing in (\ref{delS1}), and imposing the principle of least action $\delta S_{f(T)}\equiv 0$ and multiplying by $e^{-1}e^{a}_{\;\;\nu}/2$, we have the following equation of motion
\begin{eqnarray}
\frac{1}{2}f\left(e^{-1}e^{a}_{\;\;\nu}\frac{\partial e}{\partial e^{a}_{\;\;\sigma}}\right)+\frac{1}{2}f_Te^{a}_{\;\;\nu}\frac{\partial T}{\partial e^{a}_{\;\;\sigma}}-\frac{1}{2}e^{-1}e^{a}_{\;\;\nu}\partial_{\alpha}\left[ef_T\frac{\partial T}{\partial (\partial_{\alpha}e^{a}_{\;\;\sigma})}\right]-
\kappa^2 \Theta_{\nu}^{\;\;\sigma}=0\label{eqm0}\,.
\end{eqnarray}
Substituting the derivatives \cite{aldrovandi}
\begin{eqnarray}
&&\frac{\partial e}{\partial e^a_{\;\;\sigma}}= e\,e_a^{\;\;\sigma}\;,\;\frac{\partial T}{\partial e^a_{\;\;\sigma}}=-4 e_a^{\;\;\lambda} T^{\alpha}_{\;\;\nu\lambda} S_\alpha^{\;\;\nu\sigma}\;,\;\frac{\partial T}{\partial(\partial_\alpha e^a_{\;\;\sigma})}=4e_a^{\;\;\lambda}S_\lambda^{\;\;\alpha\sigma}\;,
\end{eqnarray}
in (\ref{eqm0}) we finally have the equations of motion of the $f(T)$ Gravity
\begin{eqnarray}
\frac{1}{2}f\delta^{\sigma}_{\nu}-2f_TT^{\alpha}_{\;\;\beta\nu}S_{\alpha}^{\;\;\beta\sigma}-2e^{-1}e^{a}_{\;\;\nu}\partial_{\alpha}\left[ef_Te_{a}^{\;\;\beta}S_{\beta}^{\;\;\alpha\sigma}\right]-\kappa^2\Theta_{\nu}^{\;\;\sigma}=0\label{eqmfT}\,.
\end{eqnarray}

Now we make use of identity \cite{aldrovandi}
\begin{eqnarray}
\left[e^{-1}e^{a}_{\;\;\nu}\partial_{\alpha}\left(ee_{a}^{\;\;\beta}S_{\beta}^{\;\;\alpha\sigma}\right)+T^{\alpha}_{\;\;\beta\nu}S_{\alpha}^{\;\;\beta\sigma}\right]=-\frac{1}{2}\left[G_{\nu}^{\;\;\sigma}-\frac{1}{2}\delta^{\sigma}_{\nu}T\right]\,,\label{id}
\end{eqnarray}
with $G_{\nu}^{\;\;\sigma}$ being the mixed components of the Einstein tensor, for rewrite (\ref{eqmfT}) as
\begin{eqnarray}
-2S_{\nu}^{\;\;\alpha\sigma}\partial_{\alpha}f_T+f_T G_{\nu}^{\;\;\mu}+\frac{1}{2}\delta^{\sigma}_{\nu}\left[f-f_T T\right]=\kappa^2 \Theta_{\nu}^{\;\;\sigma}\label{eqmfT2}\,.
\end{eqnarray}

This theory falls on Einstein's General Relativity with a cosmological constant, when we make $f(T)=T-2\Lambda$. Here it becomes clear that if we do not consider the sign $(-)$ in front of the matter term in action (\ref{fTaction}) in the theory, we do not return to GR for a linear $f(T)$ function, reaching a opposite signal to Einstein's equation. This fact will be crucial to show later that an invariant theory by local Lorentz transformations, as the $f(\bar{R})$ Gravity, can not fall in $f(T)$ Gravity, since these have opposite coupling signs to the matter term. 
\par 
Sotiriou et al \cite{sotiriou} have shown that $f(T)$ Gravity does not preserve its equations of motion invariant by local Lorentz transformations. It is in relation to this problem that we then construct a generalization of the Teleparallel Theory that preserves the invariant  of the equations of motion for a local Lorentz transformation. This will be addressed in the next section.

\section{Equations of motion on Generalized Teleparallel Theory}\label{sec3}
An important identity is given by $\bar{R}=-T-2\bar{\nabla}^{\mu}T^{\beta}_{\;\;\mu\beta}$, where $\bar{R}$ is the curvature scalar associated with a Riemann tensor defining solely by Levi-Civita connection $\bar{\Gamma}^{\alpha}_{\;\;\mu\nu}$, where the indices $(\mu\nu)$ are symmetric, and the covariant derivative $\bar{\nabla}$  is defined by this connection. The curvature scalar is by definition invariant through a local Lorentz transformation, but it is also invariant through a general coordinate transformation. So it would be interesting to develop a theory that generalize the TT but that the functional action depends on an invariant under local Lorentz transformations. This is not the case on $f(T)$ Gravity.
\par
We propose the following action
\begin{eqnarray}
S_{GTT}=\int d^4x\left[\frac{e}{2\kappa^2}f(\mathcal{T})+\mathcal{L}_{matter}\right]\,,\label{SGTT}
\end{eqnarray} 
where we define
\begin{eqnarray}
\mathcal{T}=-T-2a_1\bar{\nabla}^\mu T^\beta_{\;\;\;\mu\beta}=-T-2a_1\,e^{-1}\partial_\mu\left(eg^{\mu\lambda}T^{\alpha}_{\;\;\lambda\alpha}\right)\label{Tc}\,.
\end{eqnarray}
This action generalizes TT and falls on a modified $f(T)$ Gravity as well as $f(\bar{R})$ Gravity. We can show this by making the limit $a_1\rightarrow 0$, where we have $\mathcal{T}\rightarrow -T$, ergo $f(\mathcal{T})\rightarrow f(-T)$, $f_{\mathcal{T}}\rightarrow -f_T$ and the theory must be equivalent to a modified $f(T)$ (we'll see this later). Moreover, we can regain $f(\bar{R})$ Gravity, making the limit $a_1\rightarrow 1$, where we have $\mathcal{T}\rightarrow\bar{R}$, then the theory must be equivalent to $f(\bar{R})$. We show this explicitly through the equations of motion later on.
\par 
By performing the functional variation of the action (\ref{SGTT}) we obtain:
\begin{eqnarray}
\delta S_{GTT}=\frac{1}{2\kappa^2}\int d^4x\left[f\delta e+e\delta f+2\kappa^2\delta\mathcal{L}_{matter}\right]\,.\label{SGTT2}
\end{eqnarray}
As $S_{GTT}\equiv S_{GTT}\left[e^a_{\;\;\sigma}, \partial_\alpha e^a_{\;\;\sigma}, \Phi^A\right]$, in which $\Phi^A$ are the matter fields, doing,
\begin{eqnarray}
\delta S_{GTT}=\delta S_\mathcal{T}+\delta S_{matter}\,,\label{SGTT3}
\end{eqnarray}
with $\delta S_{matter}=\int d^4x\;\delta\mathcal{L}_{matter}$ in the same manner as in $f(T)$ Gravity. The functional variation of the matters term (\ref{SGTT3}) is exactly the same as given in (\ref{delSm1}).
\par 
The geometric part is
\begin{eqnarray}
\delta S_{\mathcal{T}}&=&\frac{1}{2\kappa^2}\int d^4x\left[f\;\delta e+e\;\delta f\right]\nonumber\\
&=&\frac{1}{2\kappa^2}\int d^4x \left[f\;\frac{\partial e}{\partial e^a_{\;\;\sigma}}\delta e^a_{\;\;\sigma}+e\;f_{\mathcal{T}}\delta{\mathcal{T}}\right]\,,\label{delS3}
\end{eqnarray}
where we use $f_{\mathcal{T}}=df/d\mathcal{T}$. The first term in (\ref{delS3}) is already known, we will pay attention to the second term. Performing the functional variation to $\mathcal{T}$ in (\ref{Tc}) we obtain
\begin{eqnarray}
\delta\mathcal{T}&=&-\delta T-2a_1\delta\left[e^{-1}\partial_{\mu}\left(e\,g^{\mu\beta}T^\alpha_{\;\;\beta\alpha}\right)\right]\nonumber\\
&=&-\delta T-2a_1\left[-e^{-2}\partial_\mu\left(eg^{\mu\beta}T^\alpha_{\;\;\beta\alpha}\right)\delta e+e^{-1}\delta\partial_\mu\left(eg^{\mu\beta}T^\alpha_{\;\;\beta\alpha}\right)\right]\,.
\end{eqnarray}
replacing in (\ref{delS3}) taking into account the functional variation of $T$ and $e$ we have, 
\begin{eqnarray}
&&\delta S_{\mathcal{T}}=\frac{1}{2\kappa^2}\int d^4x\bigg{\lbrace}f\frac{\partial e}{\partial e^a_{\;\;\sigma}}\delta e^a_{\;\;\sigma}-ef_{\mathcal{T}}\left[\frac{\partial T}{\partial e^a_{\;\;\sigma}}\delta e^a_{\;\;\sigma}+\frac{\partial T}{\partial(\partial_\alpha e^a_{\;\;\sigma})}\delta(\partial_\alpha e^a_{\;\;\sigma})\right]\nonumber\\
&&+2a_1\left[e^{-1}f_{\mathcal{T}}\partial_\mu(eg^{\mu\beta}T^\alpha_{\;\;\beta\alpha})\frac{\partial e}{\partial e^a_{\;\;\sigma}}\delta e^a_{\;\;\sigma}-f_{\mathcal{T}}\delta\partial_\mu(eg^{\mu\beta}T^\nu_{\;\;\beta\nu})\right]\bigg{\rbrace}\,.\label{delSTc1}
\end{eqnarray}
Now we do the integration by part in the terms containing $\delta(\partial_\alpha e^a_{\;\;\sigma})$ and $\delta\partial_\mu(eg^{\mu\beta}T^\nu_{\;\;\beta\nu})$.  
The first integration by parts is given by
\begin{eqnarray}
-\frac{1}{2\kappa^2}\int d^4x\,ef_{\mathcal{T}}\frac{\partial T}{\partial(\partial_\alpha e^a_{\;\;\sigma})}\delta(\partial_\alpha e^a_{\;\;\sigma})&=&-\frac{1}{2\kappa^2}\int d^4x\partial_\alpha\left[ef_{\mathcal{T}}\frac{\partial T}{\partial(\partial_\alpha e^a_{\;\;\sigma})}\delta e^a_{\;\;\sigma}\right]+\frac{1}{2\kappa^2}\int d^4x\partial_\alpha\left[ef_{\mathcal{T}}\frac{\partial T}{\partial(\partial_\alpha e^a_{\;\;\sigma})}\right]\delta e^a_{\;\;\sigma}\,,\nonumber\\\label{int1}
\end{eqnarray}
where the first term is zero because it is a surface term, which we consider $\delta e^{a}_{\;\;\sigma}\big|_{surface}\equiv 0$.  The second integration by parts is given by
\begin{eqnarray}
-\frac{2a_1}{2\kappa^2}\int d^4x\,f_{\mathcal{T}}\delta\partial_\mu(eg^{\mu\beta}T^\nu_{\;\;\beta\nu})&=&-\frac{2a_1}{2\kappa^2}\int d^4x\partial_\mu\left[f_{\mathcal{T}}\,\delta(eg^{\mu\beta}T^\nu_{\;\;\beta\nu})\right]+\frac{2a_1}{2\kappa^2}\int d^4x(\partial_\mu f_{\mathcal{T}})\,\delta(eg^{\mu\beta}T^\nu_{\;\;\beta\nu})\,,\label{sterm}
\end{eqnarray}
with the first term is null for being a surface term. Then we have
\begin{eqnarray}
-\frac{2a_1}{2\kappa^2}\int d^4x\,f_{\mathcal{T}}\delta\partial_\mu(eg^{\mu\beta}T^\nu_{\;\;\beta\nu})=\frac{2a_1}{2\kappa^2}\int d^4x(\partial_\mu f_{\mathcal{T}})\left[g^{\mu\beta}T^{\nu}_{\;\;\beta\nu}\frac{\partial e}{\partial e^a_{\;\;\sigma}}\delta e^a_{\;\;\sigma}+e\,T^\nu_{\;\;\beta\nu}\delta g^{\mu\beta}+eg^{\mu\beta}\delta T^\nu_{\;\;\beta\nu}\right]\;.\label{int2}
\end{eqnarray}
Making use of the following relationship
\begin{eqnarray}
\delta g^{\mu\beta}=\delta \left(\eta^{ab}e_{a}^{\;\;\mu}e_{b}^{\;\;\beta}\right)=-(g^{\beta\sigma} e_a^{\;\;\mu}\delta e^a_{\;\;\sigma}+g^{\mu\sigma}e_a^{\;\;\beta}\delta e^a_{\;\;\sigma})\,.
\end{eqnarray}
and replacing (\ref{int1}) and (\ref{int2}) in (\ref{delSTc1}), developing the terms of $\delta T^{\nu}_{\;\;\beta\nu}$ we have
\begin{eqnarray}
&&\delta S_{\mathcal{T}}=\frac{1}{2\kappa^2}\int d^4x\bigg{\lbrace}f\frac{\partial e}{\partial e^a_{\;\;\sigma}}\delta e^a_{\;\;\sigma}-e\,f_{\mathcal{T}}\frac{\partial T}{\partial e^a_{\;\;\sigma}}\delta e^a_{\;\;\sigma}+\partial_{\alpha}\left[e\,f_{\mathcal{T}}\frac{\partial T}{\partial(\partial_\alpha e^a_{\;\;\sigma})}\right]\delta e^a_{\;\;\sigma}\nonumber\\
&&+2a_1\bigg[e^{-1}f_{\mathcal{T}}\partial_{\mu}(e\,g^{\mu\beta}T^\nu_{\beta\nu})\frac{\partial e}{\partial e^a_{\;\;\sigma}}\delta e^a_{\;\;\sigma}+(\partial_\mu f_{\mathcal{T}})\bigg[g^{\mu\beta}T^\nu_{\;\;\beta\nu}\frac{\partial e}{\partial e^a_{\;\;\sigma}}\delta e^a_{\;\;\sigma}\nonumber\\
&&-eT^{\nu}_{\;\;\beta\nu}\left(g^{\beta\sigma} e_a^{\;\;\mu}\delta e^a_{\;\;\sigma}+g^{\mu\sigma}e_a^{\;\;\beta}\delta e^a_{\;\;\sigma}\right)+e\,g^{\mu\beta}\frac{\partial T^{\nu}_{\;\;\beta\nu}}{\partial e^a_{\;\;\sigma}}\delta e^a_{\;\;\sigma}+e\,g^{\mu\beta}\frac{\partial T^\nu_{\;\;\beta\nu}}{\partial(\partial_\alpha e^a_{\;\;\sigma})}\delta(\partial_\alpha e^a_{\;\;\sigma})\bigg]\bigg]\bigg{\rbrace}\,.\label{delSTc2}
\end{eqnarray}
At this point we see that we still have to do an integration by parts in the last term, ie, 
\begin{eqnarray}
\frac{2a_1}{2\kappa^2}\int d^4x\, (\partial_\mu f_{\mathcal{T}})\,e\,g^{\mu\beta}\frac{\partial T^\nu_{\;\;\beta\nu}}{\partial(\partial_\alpha e^a_{\;\;\sigma})}\delta(\partial_\alpha e^a_{\;\;\sigma})&=&\frac{2a_1}{2\kappa^2}\int d^4x\partial_\alpha\left[(\partial_\mu f_{\mathcal{T}})eg^{\mu\beta}\frac{\partial T^\nu_{\;\;\beta\nu}}{\partial(\partial_\alpha e^a_{\;\;\sigma})}\delta e^a_{\;\;\sigma}\right]\nonumber\\
&&-\frac{2a_1}{2\kappa^2}\int d^4x\partial_\alpha\left[(\partial_\mu f_{\mathcal{T}})\,e\,g^{\mu\beta}\frac{\partial T^\nu_{\;\;\beta\nu}}{\partial(\partial_\alpha e^a_{\;\;\sigma})}\right]\delta e^a_{\;\;\sigma}\,,\nonumber\\
\end{eqnarray}
where once again the first term vanishes due to be a surface term. Replacing this result in (\ref{delSTc2}) we obtain
\begin{eqnarray}
&&\delta S_{\mathcal{T}}=\frac{1}{2\kappa^2}\int d^4x\bigg{\lbrace}f\frac{\partial e}{\partial e^a_{\;\;\sigma}}-e\,f_{\mathcal{T}}\frac{\partial T}{\partial e^a_{\;\;\sigma}}+\partial_{\alpha}\left[e\,f_{\mathcal{T}}\frac{\partial T}{\partial(\partial_\alpha e^a_{\;\;\sigma})}\right]+2a_1\bigg[e^{-1}f_{\mathcal{T}}\partial_{\mu}(e\,g^{\mu\beta}T^\nu_{\beta\nu})\frac{\partial e}{\partial e^a_{\;\;\sigma}}\nonumber\\
&&+(\partial_\mu f_{\mathcal{T}})\bigg[g^{\mu\beta}T^\nu_{\;\;\beta\nu}\frac{\partial e}{\partial e^a_{\;\;\sigma}}-eT^{\nu}_{\;\;\beta\nu}\left(g^{\beta\sigma} e_a^{\;\;\mu}+g^{\mu\sigma}e_a^{\;\;\beta}\right)+e\,g^{\mu\beta}\frac{\partial T^{\nu}_{\;\;\beta\nu}}{\partial e^a_{\;\;\sigma}}\bigg]-\partial_\alpha\left[(\partial_\mu f_{\mathcal{T}})\,e\,g^{\mu\beta}\frac{\partial T^\nu_{\;\;\beta\nu}}{\partial(\partial_\alpha e^a_{\;\;\sigma})}\right]\bigg]\bigg{\rbrace}\delta e^a_{\;\;\sigma}\,.\label{delSTc3}
\end{eqnarray}
Now we must replace those derived from $T$, $e$ and $T^\nu_{\;\;\beta\nu}$ in relation tetrads and its derivatives. Taking into account the results of $ f (T) $, we have the following derivative, 
\begin{eqnarray}
&&\frac{\partial e}{\partial e^a_{\;\;\sigma}}= e\,e_a^{\;\;\sigma}\;,\;\frac{\partial T}{\partial e^a_{\;\;\sigma}}=-4 e_a^{\;\;\lambda} T^{\alpha}_{\;\;\nu\lambda} S_\alpha^{\;\;\nu\sigma}\;,\;\frac{\partial T}{\partial(\partial_\alpha e^a_{\;\;\sigma})}=4e_a^{\;\;\lambda}S_\lambda^{\;\;\alpha\sigma}\;,\;
\frac{\partial T^\nu_{\;\;\beta\nu}}{\partial e^a_{\;\;\sigma}}=-e_a^{\;\;\nu}T^\sigma_{\;\;\beta\nu}\;,\\
&&\frac{\partial T^\nu_{\;\;\beta\nu}}{\partial(\partial_\alpha e^a_{\;\;\sigma})}=e_b^{\;\;\nu}\delta_a^b\left(\delta_\beta^\alpha\delta_\nu^\sigma-\delta_\nu^\alpha\delta_\beta^\sigma\right)\,.
\end{eqnarray}
Substituting the above derivatives in (\ref{delSTc3}), making $\delta S_{GTT}\equiv 0$ in (\ref{SGTT3}) and multiplying by $\frac{1}{2}e^{-1}e^a_{\;\;\omega}$ we have the following equation of motion for the Generalized Teleparallel Theory
\begin{eqnarray}
&&\frac{1}{2}\delta^\sigma_\omega f+2f_{\mathcal{T}}T^\beta_{\;\;\nu\omega}S_\beta^{\;\;\nu\sigma}+2e^{-1}e^a_{\;\;\omega}\partial_\alpha\left[e f_{\mathcal{T}}e_a^{\;\;\beta}S_\beta^{\;\;\alpha\sigma}\right]+a_1\bigg{\lbrace}e^{-1}f_{\mathcal{T}}\delta^\sigma_\omega \partial_\mu(eg^{\mu\beta}T^\nu_{\;\;\beta\nu})+(\partial_\mu f_{\mathcal{T}})\big[\delta^\sigma_\omega g^{\mu\beta}T^\nu_{\;\;\beta\nu}\nonumber\\
&&-\left(\delta^\mu_\omega g^{\beta\sigma}T^{\nu}_{\;\;\beta\nu}+g^{\mu\sigma}T^\nu_{\;\;\omega\nu}\right)-g^{\mu\beta}T^{\sigma}_{\;\;\beta\omega}\big]-e^{-1}e^a_{\;\;\omega}\partial_\alpha\big[e(\partial_\mu f_{\mathcal{T}})(g^{\mu\alpha}e_a^{\;\;\sigma}-g^{\mu\sigma}e_a^{\;\;\alpha})\big]\bigg{\rbrace}+\kappa^2\Theta_{\omega}^{\;\;\sigma}=0\,.\label{eqmGTT}
\end{eqnarray}

Taking the limit in which $a_1\rightarrow 0$ ($\mathcal{T}\rightarrow -T,f\equiv f(-T),f_{\mathcal{T}}\rightarrow -f_T$), making $T\rightarrow -T$ the equation of motion (\ref{eqmGTT}) does not fall exactly on the equation of motion of the $f(T)$ Gravity in (\ref{eqmfT}). This happens to the fact that the relationship between the curvature scalar and the torsion scalar are through a minus sign, which prevents a theory as $f(\bar{R})$ Gravity, in which the coupling signal with matter is positive, fall in a theory like $f(T)$ Gravity, in which the coupling signal with the matter should be negative so that it falls within the GR. In the next section we'll show the equivalence between the GTT and $f(\bar{R})$ Gravity.

\section{Equivalence between GTT and $f(\bar{R})$ Gravity}\label{sec4}

Let's start this section showing the equivalence of GTT with $f(\bar{R})$ Gravity in the limit $a_1\rightarrow 1$, to general tetrads. 
\par 
Let us first establish some necessary identities, as arising from the condition of metricity
\begin{eqnarray}
&&\bar{\nabla}_{\alpha}g_{\mu\nu}=\bar{\nabla}_{\alpha}g^{\mu\nu}\equiv 0\,,\,\partial_{\alpha}g_{\mu\nu}=\bar{\Gamma}^{\lambda}_{\;\;\alpha\mu}g_{\lambda\nu}+\bar{\Gamma}^{\lambda}_{\;\;\alpha\nu}g_{\lambda\mu}\,,\,\partial_{\alpha}g^{\mu\nu}=-\bar{\Gamma}^{\mu}_{\;\;\lambda\alpha}g^{\lambda\nu}-\bar{\Gamma}^{\nu}_{\;\;\lambda\alpha}g^{\lambda\mu}\label{divg}\,.
\end{eqnarray}
With it the identity $\partial_{\alpha}e=eg^{\mu\nu}\partial_{\alpha}g_{\mu\nu}$ becomes
\begin{eqnarray}
\partial_{\alpha}e=2e\bar{\Gamma}^{\nu}_{\;\;\alpha\nu}\label{dive}\,.
\end{eqnarray} 

Now we can divide the equation of motion \eqref{eqmGTT} in terms such as
\begin{eqnarray}
&&T^{(1)}+T^{(2)}+T^{(3)}+\kappa^2\Theta^{\sigma}_{\;\;\omega}=0\label{GTT0}\,,\\
&&T^{(1)}=\frac{1}{2}\delta^\sigma_\omega f+2f_{\mathcal{T}}T^\beta_{\;\;\nu\omega}S_\beta^{\;\;\nu\sigma}+2e^{-1}e^a_{\;\;\omega}\partial_\alpha\left[e f_{\mathcal{T}}e_a^{\;\;\beta}S_\beta^{\;\;\alpha\sigma}\right]+a_1e^{-1}f_{\mathcal{T}}\delta^\sigma_\omega \partial_\mu(eg^{\mu\beta}T^\nu_{\;\;\beta\nu})\,,\label{T1}\\
&&T^{(2)}=a_1(\partial_\mu f_{\mathcal{T}})\big[\delta^\sigma_\omega g^{\mu\beta}T^\nu_{\;\;\beta\nu}-\delta^\mu_\omega g^{\beta\sigma}T^{\nu}_{\;\;\beta\nu}-g^{\mu\sigma}T^\nu_{\;\;\omega\nu}-g^{\mu\beta}T^{\sigma}_{\;\;\beta\omega}\big]\,,\label{T2}\\
&&T^{(3)}=-a_1e^{-1}e^a_{\;\;\omega}\partial_\alpha\big[e(\partial_\mu f_{\mathcal{T}})(g^{\mu\alpha}e_a^{\;\;\sigma}-g^{\mu\sigma}e_a^{\;\;\alpha})\big]\,.\label{T3}
\end{eqnarray}
Developing the last term we have
\begin{eqnarray}
&&T^{(3)}=-a_1\delta^{\sigma}_{\omega}\bar{\square} f_{\mathcal{T}}+a_1g^{\mu\sigma}\bar{\nabla}_{\omega}\bar{\nabla}_{\mu}f_{\mathcal{T}}-a_1g^{\nu\sigma}\bar{\Gamma}^{\mu}_{\;\;\omega\nu}\partial_{\mu}f_{\mathcal{T}}-a_1e^{-1}e^{a}_{\;\;\omega}\left(\partial_{\mu}f_{\mathcal{T}}\right)\Bigg[(g^{\mu\alpha}e_a^{\;\;\sigma}-g^{\mu\sigma}e_a^{\;\;\alpha})\partial_{\alpha}e+e(e_{a}^{\;\;\sigma}\partial_{\alpha}g^{\mu\alpha}\nonumber\\
&&+g^{\mu\alpha}\partial_{\alpha}e_{a}^{\;\;\sigma}-e_{a}^{\;\;\alpha}\partial_{\alpha}g^{\mu\sigma}-g^{\mu\sigma}\partial_{\alpha}e_{a}^{\;\;\alpha})\Bigg]\,.\label{T31}
\end{eqnarray}
Using \eqref{WC}, \eqref{tor}, \eqref{cont}, \eqref{conecr}, \eqref{divg} and \eqref{dive} in \eqref{T2} and \eqref{T31} we have the sum of terms $T^{(2)}$ and $T^{(3)}$ results in 
\begin{eqnarray}
&&T^{(2)}+T^{(3)}=-a_1\delta^{\sigma}_{\omega}\bar{\square} f_{\mathcal{T}}+a_1g^{\mu\sigma}\bar{\nabla}_{\omega}\bar{\nabla}_{\mu}f_{\mathcal{T}}\,.\label{T23}
\end{eqnarray}
Now we use the identity \eqref{id} in \eqref{T1}, then we can rewrite the equation of motion \eqref{GTT0}, using \eqref{T23}, as follows
\begin{eqnarray}
-f_{\mathcal{T}}G^{\mu}_{\;\;\nu}-a_1\left[\delta^{\mu}_{\nu}\bar{\square}-g^{\mu\alpha}\bar{\nabla}_{\nu}\bar{\nabla}_{\alpha}\right]f_{\mathcal{T}}+\frac{1}{2}\left[-\mathcal{T}f_{\mathcal{T}}+f\right]\delta^{\mu}_{\nu}+2S_{\nu}^{\;\;\alpha\mu}\partial_{\alpha}f_{\mathcal{T}}+\kappa^2\Theta^{\mu}_{\;\;\nu}=0\label{GTT1}\,.
\end{eqnarray}
Considering $\mathcal{T}\equiv \mathcal{T}(-T-a_1B)$, with $B=2\partial_\mu(eg^{\mu\beta}T^\nu_{\;\;\beta\nu})$, we have to GTT will only be equivalent to $f(\bar{R})$ Gravity in the limit $a_1\rightarrow 1$, so $\mathcal{T}\rightarrow \bar{R}$ and the term $2S_{\nu}^{\;\;\alpha\mu}\partial_{\alpha}f_{\mathcal{T}}$ must be identically zero, as shown in Section III subsection C to \cite{bahamonde}. When this term vanishes, we have exactly one theory invariant by local Lorentz transformations, which occurs only when $a_1\rightarrow 1$ and thus the equation \eqref{GTT1} becomes identical to the $f(\bar{R})$ Gravity, which is covariant and  independent of the chosen of set of tetrads.
\par
In the next section we will specify a set of tetrads that explicitly show the equivalence between the two theories to the limit referred to above.

\subsection{Friedmann-Lemaitre-Robertson-Walker case}\label{subsec4.1}

In this section we explicitly show that the GTT equations of motion in (\ref{eqmGTT}), are exactly the same as $f(\bar{R})$ Gravity for the particular limit in which $a_1\rightarrow 1$. We can then begin comparing the equations of motion for a easier symmetry of the metric, as the maximum symmetry for the cosmological Friedmann-Lemaitre-Robertson-Walker (FLRW) flat metric
\begin{eqnarray}
dS^2_{FLRW}=dt^2-a^2(t)\left(dx^2+dy^2+dz^2\right)\label{eleFLRW}\;.
\end{eqnarray}
Considering now the case of cosmology, with line element FLRW flat (\ref{eleFLRW}), for a diagonal tetrad $[e^{a}_{\;\;\sigma}]=diag[1,a(t),a(t),a(t)]$, we have that the equations (\ref{eqmGTT}) become
\begin{eqnarray}
\kappa^2\Theta_0^{\;\;0}=\frac{1}{2a^2}\bigg{\lbrace}6a_1a\dot{a}\left(\frac{d}{dt}f_{\mathcal{T}}\right)+\left[12\left(1-a_1\right)(\dot{a})^2-6a_1a\ddot{a}\right]f_{\mathcal{T}}-fa^2\bigg{\rbrace}\;,\label{eqmGTTFLRW00}
\end{eqnarray}
\begin{eqnarray}
\kappa^2\Theta_1^{\;\;1}=\kappa^2\Theta_2^{\;\;2}=\kappa^2\Theta_3^{\;\;3}=-\frac{1}{2a^2}\bigg{\lbrace}2a\left(a_1 a\frac{d}{dt}+2\dot{a}\right)\frac{d}{dt}f_{\mathcal{T}}+\left[\left(4-6a_1\right)a\ddot{a}+\left(8-12a_1\right)(\dot{a})^2\right]f_{\mathcal{T}}-fa^2\bigg{\rbrace} \;,\label{eqmGTTFLRW11}
\end{eqnarray}
where $\dot{a}=(d/dt)a$ and $\ddot{a}=(d^2/dt^2)a$.
\par 
We can now compare these equations with those obtained from the $f(\bar{R})$ Gravity, whose equations of motion are \cite{fR}
\begin{eqnarray}
\kappa^2\Theta^{\mu}_{\;\;\nu}=f_{\bar{R}}\bar{R}^{\mu}_{\;\;\nu}-\frac{1}{2}\delta^{\mu}_{\nu}f+\left(\delta^{\mu}_{\nu}\bar{\square} -g^{\mu\beta}\bar{\nabla}_{\beta}\bar{\nabla}_{\nu}\right)f_{\bar{R}}\;.\label{eqmfR}
\end{eqnarray}
Considering the flat FLRW metric (\ref{eleFLRW}), the equations (\ref{eqmfR}) provide us
\begin{eqnarray}
\kappa^2\Theta_0^{\;\;0}=\frac{1}{2a}\left[6\dot{a}\frac{d}{dt}f_R-6\ddot{a}f_R-af\right]\;, \label{eqmfRFLRW00}
\end{eqnarray}
\begin{eqnarray}
\kappa^2\Theta_1^{\;\;1}=\kappa^2\Theta_2^{\;\;2}=\kappa^2\Theta_3^{\;\;3}=-\frac{1}{2a^2}\left[\left(2a^2\frac{d}{dt}+4a\dot{a}\right)\frac{d}{dt}f_R-\left(2a\ddot{a}+4(\dot{a})^2\right)f_R-fa^2\right]\;.\label{eqmfRFLRW11}
\end{eqnarray}
Subtracting (\ref{eqmGTTFLRW00}) from (\ref{eqmfRFLRW00}) we have 
\begin{eqnarray}
0=\frac{3}{a^2}\left\{2(1-a_1)\dot{a}^2f_{\mathcal{T}}+a\dot{a}\left(a_1\frac{df_{\mathcal{T}}}{dt}-\frac{df_{\bar{R}}}{dt}\right)+a\ddot{a}(f_{\bar{R}}-a_1f_{\mathcal{T}})+\frac{a^2}{6}\left[f(\bar{R})-f(\mathcal{T})\right]\right\}\;.\label{sub1}
\end{eqnarray}
Subtravting (\ref{eqmGTTFLRW11}) from (\ref{eqmfRFLRW11}) we obtain 
\begin{eqnarray}
0=&&\frac{1}{a^2}\Big\{a^2\left(a_1\frac{d^2f_{\mathcal{T}}}{dt^2}-\frac{d^2f_{\bar{R}}}{dt^2}\right)+2a\dot{a}\left(\frac{df_{\mathcal{T}}}{dt}-\frac{df_{\bar{R}}}{dt}\right)+a\ddot{a}[(2-3a_1)f_{\mathcal{T}}+f_{\bar{R}}]+\dot{a}^2[(4-6a_1)f_{\mathcal{T}}+2f_{\bar{R}}]\nonumber\\
&&+\frac{a^2}{2}\left[f(\bar{R})-f(\mathcal{T})\right]\Big\}\;.\label{sub2}
\end{eqnarray}
Now we see clearly that to the limit at which $a_1\rightarrow 1$, we have $\{\mathcal{T}\rightarrow \bar{R},f(\mathcal{T})\rightarrow f(\bar{R}),f_{\mathcal{T}}\rightarrow f_{\bar{R}}\}$, then (\ref{sub1}) and (\ref{sub2}) are identically null, showing the equivalence of equations of motion between GTT and $f(\bar{R})$ for this limit. The conclusion is that the GTT is only invariant under local Lorentz transformations and at the same time invariant by general coordinates transformations to the limit at which $a_1\rightarrow 1$.

\subsection{Spherically symmetric case}\label{subsec4.2}

We have demonstrated in general that the GTT is equivalent to gravity $ f (R) $, but in addition to explain this through a metric with specific symmetry, we want to leave the equations of motion open for further analysis of this theory.

\par
Let us now consider the case of a spherically symmetric and static line element
\begin{eqnarray}
dS^2=e^{a(r)}dt^2-e^{b(r)}dr^2-r^2\left(d\theta^2+\sin^2\theta d\phi^2\right)\label{eles}\;,
\end{eqnarray}
we can choose the following diagonal tetrad $[e^{a}_{\;\;\sigma}]=diag[e^{a(r)/2},e^{b(r)/2},r,r\sin\theta]$, which taking into account (\ref{eqmGTT}), provides us the following equations of motion
\begin{eqnarray}
\kappa^2\Theta_0^{\;\;0}&=&-\frac{e^{-b}}{4r^2}\bigg{\lbrace}4a_1r^2\frac{d^2}{dr^2}f_{\mathcal{T}}+(8r-2a_1r^2b')\frac{d}{dr}f_{\mathcal{T}}+\bigg[\left(a_1r^2a'+4(a_1-1)r\right)b'\nonumber\\
&&+4(a_1-1)e^b-a_1r^2\left(2a''+(a')^2\right)-4(2a_1-1)ra'-8(a_1-1)\bigg]f_{\mathcal{T}}+2\,f\,r^2e^b\bigg{\rbrace}\;, \label{diag00}
\end{eqnarray}
\begin{eqnarray}
\kappa^2\Theta_1^{\;\;1}&=&\frac{e^{-b}}{4r^2}\bigg{\lbrace}a_1\left(2r^2a'+8r\right)\frac{d}{dr}f_{\mathcal{T}}+\bigg[a_1r(ra'+4)b'+4(a_1-1)e^b\nonumber\\
&&-a_1\left(2r^2a''+r^2(a')^2\right)+8(1-a_1)ra'+8(1-a_1)\bigg]f_{\mathcal{T}}+2\,f\,r^2e^b\bigg{\rbrace}\;,\label{diag11}
\end{eqnarray}
\begin{eqnarray}
\kappa^2\Theta_2^{\;\;1}=\frac{(a_1-1)\cos\theta\frac{d}{dr}f_{\mathcal{T}}}{r^2\sin\theta}=0\;,\label{diag21}
\end{eqnarray}
\begin{eqnarray}
\kappa^2\Theta_2^{\;\;2}&=&\kappa^2\Theta_3^{\;\;3}=\frac{e^{-b}}{4r^2}\bigg{\lbrace}4a_1r^2\frac{d^2}{dr^2}f_{\mathcal{T}}-2r\left(a_1rb'-ra'-2\right)\frac{d}{dr}f_{\mathcal{T}}+\bigg[\left((a_1-1)r^2a'+(4a_1-2)r\right)b'+4a_1e^b\nonumber\\
&&+2(1-a_1)r^2a''+(1-a_1)r^2(a')^2+(6-8a_1)ra'-4(2a_1-1)\bigg]f_{\mathcal{T}}+2\,f\,r^2e^b\bigg{\rbrace}\;,\label{diag22}
\end{eqnarray}
where $'$ denotes derivation in relation to radial coordinate $r$. Taking the metric (\ref{eles}) to the equations of the $f(\bar{R})$ Gravity in (\ref{eqmfR}), we obtain
\begin{eqnarray}
\kappa^2\Theta_0^{\;\;0}&=&-\frac{e^{-b}}{4r}\bigg{\lbrace}4r\frac{d^2}{dr^2}f_{\bar{R}}+(8-2rb')\frac{d}{dr}f_{\bar{R}}+\left[r(a'b'-2a''-(a')^2)-4a'\right]f_{\bar{R}}+2fre^b\bigg{\rbrace}\;,\label{fRs00}\\
\kappa^2\Theta_1^{\;\;1}&=&\frac{e^{-b}}{4r}\bigg{\lbrace}(2ra'+8)\frac{d}{dr}f_{\bar{R}}+\left[(ra'+4)b'-2ra''-r(a')^2\right]f_{\bar{R}}+2fre^b\bigg{\rbrace}\;,\label{fRs11}\\
\kappa^2\Theta_2^{\;\;2}&=&\kappa^2\Theta_3^{\;\;3}=\frac{e^{-b}}{2r^2}\bigg{\lbrace}2r^2\frac{d^2}{dr^2}f_{\bar{R}}-r(rb'-ra'-2)\frac{d}{dr}f_{\bar{R}}+(rb'+2e^b-ra'-2)f_{\bar{R}}+fr^2e^b\bigg{\rbrace}\;.\label{fRs22}
\end{eqnarray}

Here first we noticed that if $a_1\neq 1$, exists an equation (\ref{diag21}) outside the diagonal for GTT, resulting in the restriction of functional form $f(\mathcal{T})=c_1\mathcal{T}+c_0$, com $c_0,c_1\in\Re$. Then we have the same constraint to $f(T)$ Gravity in this case \cite{daouda}.
\par 
We also see that to the limit at which $a_1\rightarrow 1$,  $\{\mathcal{T}\rightarrow \bar{R},f(\mathcal{T})\rightarrow f(\bar{R}),f_{\mathcal{T}}\rightarrow f_{\bar{R}}\}$, all equations (\ref{diag00})-(\ref{diag22}) for GTT are identical to $f(\bar{R})$ given in (\ref{fRs00})-(\ref{fRs22}).
\par
Now choose a set of non-diagonal tetrads 
\begin{eqnarray}\label{ndtetrad}
\{e^{a}_{\;\;\mu}\}=\left[\begin{array}{cccc}
e^{a/2}&0&0&0\\
0&e^{b/2}\sin\theta\cos\phi & r\cos\theta\cos\phi &-r\sin\theta\sin\phi\\
0&e^{b/2}\sin\theta\sin\phi &
r\cos\theta\sin\phi &r\sin\theta
\cos\phi  \\
0&e^{b/2}\cos\theta &-r\sin\theta  &0
\end{array}\right]\;,
\end{eqnarray}  
the equations to GTT in (\ref{eqmGTT}) provide us
\begin{eqnarray}
\kappa^2\Theta_0^{\;\;0}&=&-\frac{e^{-b}}{4r^2}\bigg{\lbrace}4a_1r^2\frac{d^2}{dr^2}f_{\mathcal{T}}-\left(2a_1r^2b'-8(a_1-1)re^{b/2}-8r\right)\frac{d}{dr}f_{\mathcal{T}}+\bigg[\left(a_1r^2a'+4(a_1-1)r\right)b'+\big(4(a_1-1)ra'\nonumber\\
&&+8(a_1-1)\big)e^{b/2}-a_1r^2\left(2a''+(a')^2\right)-(8a_1-4)ra'-8(a_1-1)\bigg]f_{\mathcal{T}}+2\,f\,r^2e^b\bigg{\rbrace}\;,\label{ndiag00}
\end{eqnarray}
\begin{eqnarray}
\kappa^2\Theta_1^{\;\;1}&=&\frac{e^{-3b/2}}{4r^2}\bigg{\lbrace}2a_1r(ra'+4)e^{b/2}\frac{d}{dr}f_{\mathcal{T}}+\bigg[a_1r(a'r+4)e^{b/2}b'+\big[4(a_1-1)ra'+8(a_1-1)\big]e^b\nonumber\\
&&-\left(a_1r^2\left(2a''+(a')^2\right)-8(1-a_1)ra'-8(1-a_1)\right)e^{b/2}\bigg]f_{\mathcal{T}}+2\,f\,r^2e^{3b/2}\bigg{\rbrace}\;,\label{ndiag11}
\end{eqnarray}
\begin{eqnarray}
\kappa^2\Theta_2^{\;\;2}&=&\kappa^2\Theta_3^{\;\;3}=\frac{e^{-b}}{4r^2}\bigg{\lbrace}4a_1r^2\frac{d^2}{dr^2}f_{\mathcal{T}}-\left(2a_1b'-4(a_1-1)re^{b/2}-2r^2a'-4r\right)\frac{d}{dr}f_{\mathcal{T}}\nonumber\\
&&+\bigg[\left((a_1-1)r^2a'+(4a_1-2)r\right)b'+4e^b+\left(4(a_1-1)ra'+8(a_1-1)\right)e^{b/2}\nonumber\\
&&-2(a_1-1)r^2a''-(a_1-1)r^2(a')^2+(6-8a_1)ra'-(8a_1-4)\bigg]f_{\mathcal{T}}+2\,f\,r^2 e^b\bigg{\rbrace}\;.\label{ndiag22}
\end{eqnarray}

We can then see that in this case the equations of motion are diagonals. But equivalence of the GTT with the $f(\bar{R})$ Gravity only gives to the limit $a_1\rightarrow 1$, when the equations (\ref{fRs00})-(\ref{fRs22}) and (\ref{ndiag00})-(\ref{ndiag22}) are identical.

\section{Equivalence between GTT and a particular case of the $f(T,B)$ Gravity}\label{sec5}

In this section we make an important observation. When we were finishing the calculation of the non-diagonal tetrads case of the previous subsection, we note that a group have submitted exactly the same idea of our work here. The so-call $f(T,B)$ Gravity \cite{bahamonde}, with $B=-2\bar{\nabla}^{\mu}T^{\nu}_{\;\;\mu\nu}$, is a more general theory that presented here, where the algebraic function contained in action, may be any analytic function of the variables $T$ and $B$. We noted then that the equivalence of this theory with the $f(\bar{R})$ Gravity is given only for the specific functional form $f(T,B)\equiv f(-T+B)=f(\bar{R})$. Compared to our theory, we have the GTT is a particular case of $f(T,B)$ Gravity, when $f(T,B)\equiv f(-T+a_1B)=f(\mathcal{T})$. We can show this again explicitly using equations of motion.
\par   
The equation of motion for $f(T,B)$ Gravity is given by
\begin{eqnarray}
&&2\delta^{\lambda}_{\nu}\bar{\square}f_B-2\bar{\nabla}^{\lambda}\bar{\nabla}_{\nu}f_B+Bf_B\delta^{\lambda}_{\nu}+4\partial_{\mu}\left(f_B+f_T\right)S_{\nu}^{\;\;\mu\lambda}\nonumber\\
&&+4e^{-1}e^{a}_{\;\;\nu}\partial_{\mu}\left(ee_{a}^{\;\;\beta}S_{\beta}^{\;\;\mu\lambda}\right)f_T-4f_TT^{\sigma}_{\;\;\mu\nu}S_{\sigma}^{\;\;\lambda\mu}-f\delta^{\lambda}_{\nu}=2\kappa^2\Theta_{\nu}^{\;\;\lambda}\label{fTB}\;.
\end{eqnarray} 

The first observation here is that this theory does not fall in $f(T)$ Gravity on general, as well as our GTT, as mentioned at the end of the section \ref{sec3}. Making $f(T,B)\equiv f(T)$, ergo $f_B=0$, the equation of motion (\ref{fTB}), using the identity (\ref{id}), becomes
\begin{eqnarray}
4(\partial_{\mu}f_T)S_{\nu}^{\;\;\mu\lambda}-2f_TG_{\nu}^{\;\;\lambda}+\delta^{\lambda}_{\nu}(Tf_T-f)=2\kappa^2\Theta_{\nu}^{\;\;\lambda}\label{fTBfT}\;.
\end{eqnarray}
This equation is not equal to (\ref{eqmfT2}) for $f(T)$ Gravity, and can not fall on GR when $f(T)\equiv T-2\Lambda$, due to sign. This shows that the $f(T,B)$ Gravity also not returns on $f(T)$ Gravity on general.
\par 
Now we can show that the particular case $f(-T+a_1B)$ this theory falls in our GTT. We take the FLRW metric (\ref{eleFLRW}) with diagonal tetrads $[e^{a}_{\;\;\mu}]=diag[1,a,a,a]$, the equations of motion (\ref{fTB}) provides us with 
\begin{eqnarray}
&&\kappa^2\Theta_0^{\;\;0}=-\frac{1}{2a^2}\bigg{\lbrace}12(\dot{a})^2f_T-6a\dot{a}\frac{d}{dt}f_B+2\left[3a\ddot{a}+6(\dot{a})^2\right]f_B+fa^2\bigg{\rbrace}\;,\label{fTBFLRW00}\\
&&\kappa^2\Theta_1^{\;\;1}=\kappa^2\Theta_2^{\;\;2}=\kappa^2\Theta_3^{\;\;3}=\frac{1}{2a^2}\bigg{\lbrace}4a(\dot{a})\left(\frac{d}{dt}f_T\right)+\left[4a\ddot{a}+8(\dot{a})^2\right]f_T-2a^2\left(\frac{d^2}{dt^2}f_B\right)\nonumber\\
&&+\left[6a\ddot{a}+12(\dot{a})^2\right]f_B+fa^2\bigg{\rbrace}\;.\label{fTBFLRW11}
\end{eqnarray}
Now identifing $f(-T+a_1B)=f(\mathcal{T})$, recall that $\mathcal{T}$ is given in (\ref{Tc}), then
\begin{eqnarray}
f_T=\frac{\partial f}{\partial T}=\frac{\partial \mathcal{T}}{\partial T}\frac{df}{d\mathcal{T}}=-\frac{df}{d\mathcal{T}}\;,\;f_B=\frac{\partial f}{\partial B}=\frac{\partial \mathcal{T}}{\partial B}\frac{df}{d\mathcal{T}}=a_1\frac{df}{d\mathcal{T}}\;,\label{id2}
\end{eqnarray}
We have that the equations (\ref{fTBFLRW00}) and (\ref{fTBFLRW11}) are identical from the GTT (\ref{eqmGTTFLRW00}) and (\ref{eqmGTTFLRW11}), thus showing the equivalence between the theories.
\par 
We can also confirm this by choosing the spherical symmetry for the metric (\ref{eles}), first for diagonal tetrads $[e^{a}_{\;\;\mu}]=diag[e^{a/2},e^{b/2},r,r\sin\theta]$, ergo, the equations (\ref{fTB}) provide us  
\begin{eqnarray}
&&\kappa^2\Theta_0^{\;\;0}=\frac{e^{-b}}{4r^2}\bigg{\lbrace}8r\frac{d}{dr}f_T-2\left[2rb'+2e^b-2ra'-4\right]f_T-2r^2\left[2\frac{d}{dr}-b'\right]\frac{d}{dr}f_B\nonumber\\
&&+\left[-r(ra'+4)b'-4e^b+r\left(2ra''+r(a')^2+8a'\right)+8\right]f_B-2fr^2e^b\bigg{\rbrace}\;,\label{fTBdiag00} \\
&&\kappa^2\Theta_1^{\;\;1}=\frac{e^{-b}}{4r^2}\bigg{\lbrace}2\left[2e^b-4ra'-4\right]f_T+2r\left[ra'+4\right]\frac{d}{dr}f_B\nonumber\\
&&+\left[r\left(ra'+4\right)b'+4e^b-r\left(2ra''-r(a')^2-8a'\right)-8\right]f_B+2r^2fe^b\bigg{\rbrace}\;,\label{fTBdiag11} \\
&&\kappa^2\Theta_2^{\;\;1}=-\frac{1}{r^2\sin\theta}\left[\cos\theta\frac{d}{dr}f_T+\cos\theta\frac{d}{dr}f_B\right]=0\;,\label{fTBdiag21} \\
&&\kappa^2\Theta_2^{\;\;2}=\kappa^2\Theta_3^{\;\;3}=\frac{e^{-b}}{4r^2}\bigg{\lbrace}2r(ra'+2)\frac{d}{dr}f_T+\left[-r\left(ra'+2\right)b'+r^2\left(2a''+(a')^2\right)+6ra'+4\right]f_T\nonumber\\
&&-2r^2\left[2\frac{d}{dr}-b'\right]\frac{d}{dr}f_B+\left[-r(ra'-4)b'-4e^b+r^2\left(2a''+(a')^2\right)+8ra'+8\right]f_B-2fr^2e^b\bigg{\rbrace}\;.\label{fTBdiag22}
\end{eqnarray}
Again we have the equivalence of the equations of motion (\ref{fTBdiag00})-(\ref{fTBdiag22}) with (\ref{diag00})-(\ref{diag22}), for the identifications $f(T,B)=f(\mathcal{T})$ and (\ref{id2}).
\par
By taking the choice of non-diagonal tetrads(\ref{ndtetrad}), the equations of motion from the $f(T.B)$ Gravity (\ref{fTB}) provide us
\begin{eqnarray}
&&\kappa^2\Theta_0^{\;\;0}=-\frac{e^{-5b/2}}{4r^2}\bigg{\lbrace}8r\left[e^{2b}-e^{3b/2}\right]\frac{d}{dr}f_T+\left[4re^{3b/2}b'+2(2ra'+4)e^{2b}-4(ra'+2)e^{3b/2}\right]f_T\nonumber\\
&&+\left[4r^2e^{3b/2}\frac{d}{dr}+\left(8re^{2b}-2r^2e^{3b/2}\right)\right]\frac{d}{dr}f_B\nonumber\\
&&+\left[r\left(ra'+4\right)e^{3b/2}b'+4\left(ra'+2\right)e^{2b}-\left(2r^2a''+r(a')^2+8ra'+8\right)e^{3b/2}\right]f_B+2fr^2e^{5b/2}\bigg{\rbrace}\;,\label{fTBndiag00}
\end{eqnarray}
\begin{eqnarray}
&&\kappa^2\Theta_1^{\;\;1}=\frac{e^{-5b/2}}{4r^2}\bigg{\lbrace}\left[4(ra'+2)e^{2b}-8(ra'+1)e^{3b/2}\right]f_T+2r(ra'+4)e^{3b/2}\frac{d}{dr}f_B\nonumber\\
&&+\left[r(ra'+4)e^{3b/2}b'+4(ra'+2)e^{2b}-(2r^2a''+r^2(a')^2+8ra'+8)e^{3b/2}\right]f_B+2fr^2e^{5b/2}\bigg{\rbrace}\;,\label{fTBndiag11}
\end{eqnarray}
\begin{eqnarray}
&&\kappa^2\Theta_2^{\;\;2}=\kappa^2\Theta_3^{\;\;3}=\frac{e^{-5b/2}}{4r^2}\bigg{\lbrace}\left[4re^{2b}-2r(ra'+2)e^{3b/2}\right]\frac{d}{dr}f_T\nonumber\\
&&+\left[r(ra'+2)e^{3b/2}b'-4e^{5b/2}+4(ra'+2)e^{2b}-\left(2r^2a''+r^2(a')^2+6ra'+4\right)e^{3b/2}\right]f_T\nonumber\\
&&+\left[4r^2e^{3b/2}\frac{d}{dr}+4re^{2b}-2r^2e^{3b/2}b'\right]\frac{d}{dr}f_B\nonumber\\
&&+\left[r(ra'+4)e^{3b/2}b'+4(ra'+2)e^{2b}-\left(2r^2a''+r^2(a')^2+8ra'+8\right)e^{3b/2}\right]f_B+2fr^2e^{5b/2}\bigg{\rbrace}\;.\label{fTBndiag22}\\
\end{eqnarray}
Just as before, making identifications $f(T,B)=f(\mathcal{T})$ and (\ref{id2}), the equations (\ref{fTBndiag00})-(\ref{fTBndiag22}) are identical from the GTT in (\ref{ndiag00})-(\ref{ndiag22}), confirming again the equivalence of these theories.

\section{Applications to GTT}\label{sec6}

\subsection{Reconstrction for de Sitter Universe}\label{subsec6.1}

A method to obtain the functional form of the algebraic function $f(\mathcal{T})$ is the so-called reconstruction. This method is to specify a model that fix the material content of the theory in terms of scalar $\mathcal{T}$, allowing to reconstruct the functional form of $f(\mathcal{T})$ through of the equations of motion of the theory. 
\par 
We will choose the particular case of flat FLRW metric in which $a(t)=a_0\exp[H_0(t-t_0)], a_0,H_0,t_0\in\Re_{+}$, it provides us with the model of de Sitter universe, where $H(t)=\dot{a}/a=H_0$. In this case, using (\ref{Tc}), we have that $H_0(\mathcal{T})=\sqrt{\mathcal{T}/[6(1-3a_1)]}$, $\dot{H}\equiv 0$ and $(d/dt)f_{\mathcal{T}}=f_{\mathcal{T}\mathcal{T}}(d/dt)\mathcal{T}\equiv 0$. Knowing that $\kappa^2\Theta_{0}^{\;\;0}=\kappa^2\rho=3H_0^2$, the equation (\ref{eqmGTTFLRW00}) provide us
\begin{eqnarray}
3[H_0(\mathcal{T})]^2=3[H_0(\mathcal{T})]^2(2-3a_1)f_{\mathcal{T}}(\mathcal{T})-\frac{1}{2}f(\mathcal{T})\label{eqr1}\,,
\end{eqnarray}
integrating with respect to that $\mathcal{T}$ results in
\begin{eqnarray}
f(\mathcal{T})=\mathcal{T}+\left[(2-3a_1)\mathcal{T}\right]^{(1-3a_1)/(2-3a_1)}c_1\;,\;c_1\in\Re\,.\label{fTc1}
\end{eqnarray}

\subsection{Spherially symmetric type-de Sitter solution}\label{subsec6.2}

We take here the limit at which $a_1\rightarrow 0$ in (\ref{eqmGTT}), that after use the identity (\ref{id}) and the consideration $\mathcal{T}\rightarrow -T$, results in 
\begin{eqnarray}
f_{T}(-T)G_{\omega}^{\;\;\sigma}+\frac{1}{2}\delta^{\sigma}_{\omega}\left[f(-T)-Tf_{T}(-T)\right]=-\kappa^2\Theta_{\omega}^{\;\;\sigma}\label{GTTfT}\;.
\end{eqnarray}
As in $f(\bar{R})$ Gravity \cite{capozziello}, we can consider a very specific case where $\bar{R}\equiv \bar{R}_0=-T_0+B_0,\mathcal{T}\equiv \mathcal{T}_0=-T_0+a_1B_0=-T_0$, with $\bar{R}_0,T_0,B_0\in \Re$ and $B_0$ is defined by (\ref{Tc}). In the case of a perfect fluid $\Theta_{\nu}^{\;\;\lambda}=diag[\rho_0,-p_0,-p_0,-p_0]$, and $\partial_{\mu}f_T=f_{TT}\partial_{\mu}T_0\equiv 0$, which results in the equations
\begin{eqnarray}
\bar{R}_{\omega}^{\;\;\sigma}=-\frac{\kappa^2}{f_{T_0}(-T_0)}\Theta_{\omega}^{\;\;\sigma}+\frac{1}{2}\delta^{\sigma}_{\omega}\left[\bar{R}_0+T_0-\frac{f(-T_0)}{f_{T_0}(-T_0)}\right]\;,\label{GTTdeSitter}
\end{eqnarray}
which taking the trace results in
\begin{eqnarray}
B_0=\frac{\kappa^2}{f_{T_0}(-T_0)}(\rho_0-3p_0)+2\frac{f(-T_0)}{f_{T_0}(-T_0)}-T_0\;.
\end{eqnarray}

Considering now the line element (\ref{eles}), for $b(r)=-a(r)$ and $p_0=-\rho_0$ (type-dark energy), we can integrate the equations of motion (\ref{GTTdeSitter}), where we get the following solution
\begin{eqnarray}
a(r)=-b(r)=\log\left[1+\frac{c_1}{2f_{T_0}(-T_0)r}+\frac{f(-T_0)-f_{T_0}(-T_0)T_0+2\kappa^2\rho_0}{6f_{T_0}(-T_0)}r^2\right]\;.\label{ssol}
\end{eqnarray}
This is a static type-de Sitter solution where we can identify effective cosmological constant $(-\Lambda_{eff}/3)=[f(-T_0)-f_{T_0}(-T_0)T_0+2\kappa^2\rho_0]/[6f_{T_0}(-T_0)]$. A solution type-de Sitter was also previously obtained in $f(\bar{R})$ Gravity for $a(r)=-b(r)$ and $\bar{R}=\bar{R}_0$ \cite{capozziello}. {\bf We emphasize here that this solution it comes to different theory to $f(T)$ Gravity, because the GTT does not fall in $f(T)$ Gravity for $a_1\rightarrow 0$, except for the special case where $f(T)$ is a odd analytic function, that is $f(-T)=-f(T)$.}

\subsection{Evolution for the state parameter of the dark energy}\label{subsec6.3}

A good test for our theory is the evolution of a model of the universe. This can discard or keep a theory depending on whether it is in agreement with the observational data.
\par 
Let's follow the procedure found in \cite{bamba2} to determine the state parameter $\omega_{DE}$. For a universe permeated by a perfect fluid, of which equation of state is governed by $p=\omega\rho$, we can rewrite the equations of motion \eqref{eqmGTTFLRW00} and \eqref{eqmGTTFLRW11} as
\begin{eqnarray}
&&3H^2=\kappa^2G_{eff}\left(\rho_{m}+\rho_{DE}\right)\,,\,H=\frac{\dot{a}}{a}\,,\,G_{eff}=\frac{1}{(3a_1-2)f_{\mathcal{T}}}\,,\,\rho_{DE}=\frac{1}{\kappa^2}\left[3a_1\left(\dot{H}f_{\mathcal{T}}-H\dot{f}_{\mathcal{T}}\right)-\frac{1}{2}f\right]\label{H2}\,,\\
&&\dot{H}=-\frac{\kappa^2}{2}G_{eff}\left(p_{m}+p_{DE}+\rho_{m}+\rho_{DE}\right)\,,\,p_{DE}=\frac{1}{\kappa^2}\left[a_1\ddot{f}_{\mathcal{T}}+H\dot{f}_{\mathcal{T}}-\frac{1}{2}f\right]\label{Hdot}\,.
\end{eqnarray}
Now we can defining the state parameter of dark energy by
\begin{eqnarray}
\omega_{DE}=\frac{p_{DE}}{\rho_{DE}}=\frac{a_1\ddot{f}_{\mathcal{T}}+H\dot{f}_{\mathcal{T}}-\frac{1}{2}f}{3a_1\left(\dot{H}f_{\mathcal{T}}-H\dot{f}_{\mathcal{T}}\right)-\frac{1}{2}f}\label{omegaDE}\,.
\end{eqnarray}

\begin{figure}[h]
\centering
\begin{tabular}{rl}
\includegraphics[height=5cm,width=8cm]{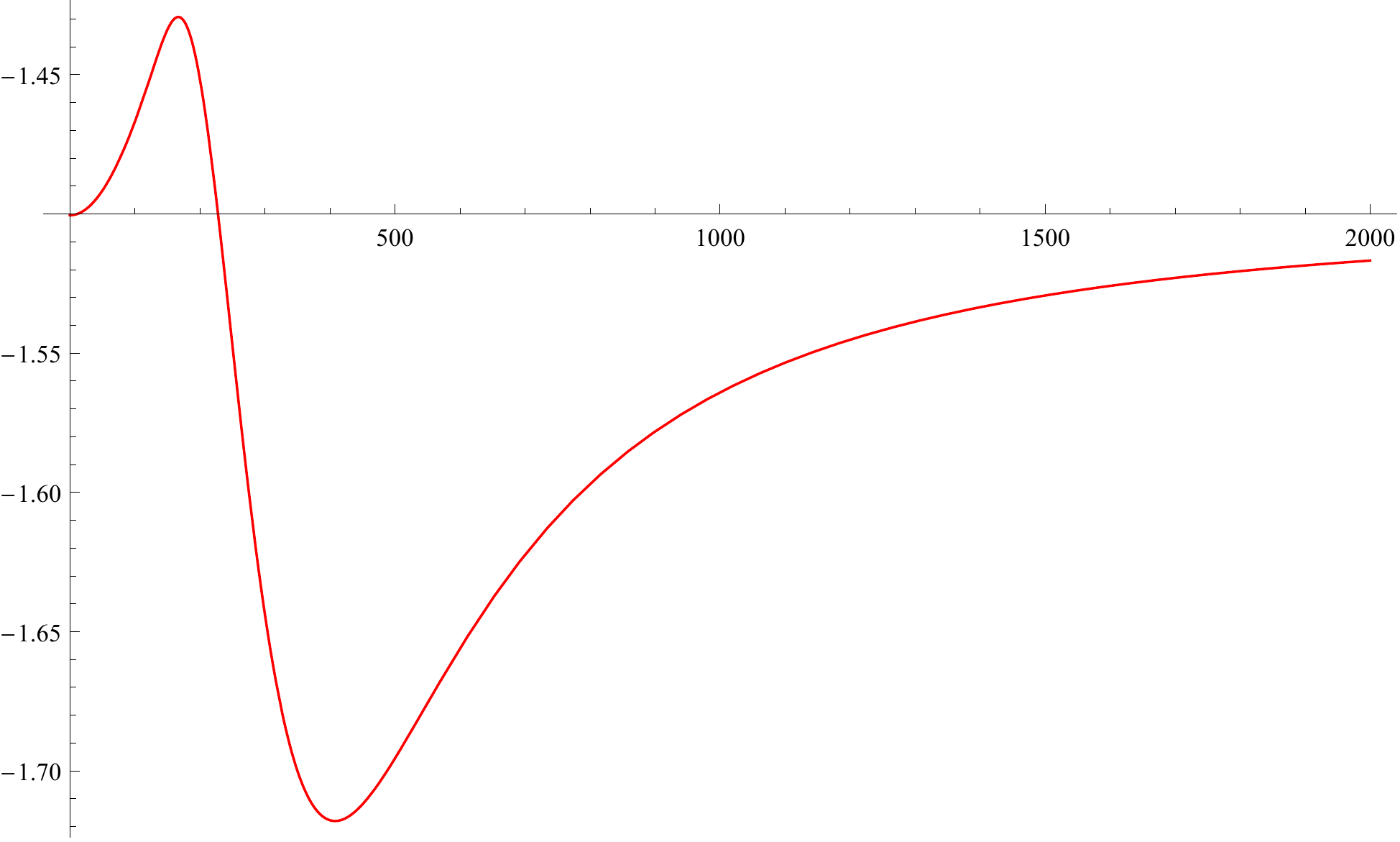}&
\includegraphics[height=5cm,width=8cm]{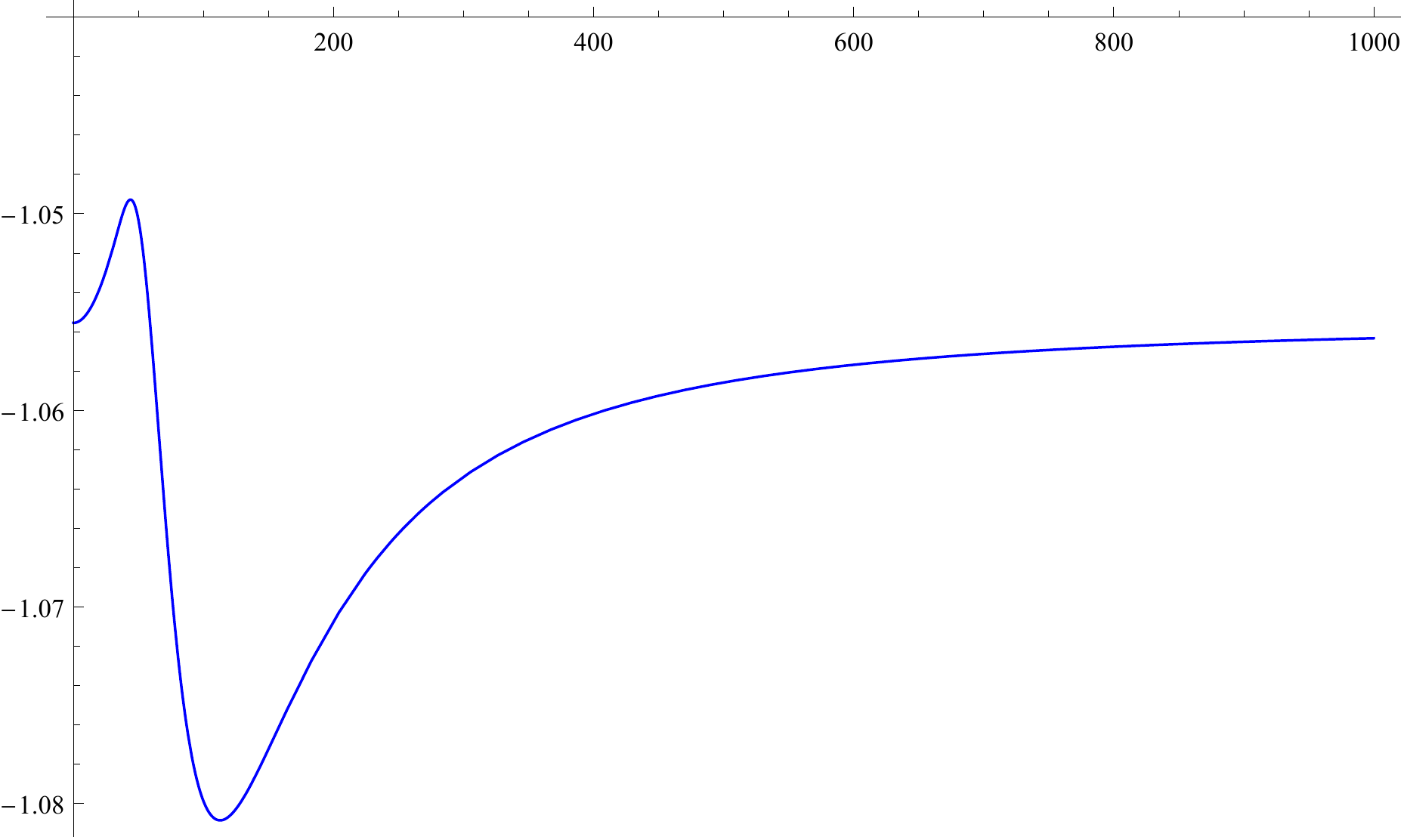}\\
\includegraphics[height=5cm,width=8cm]{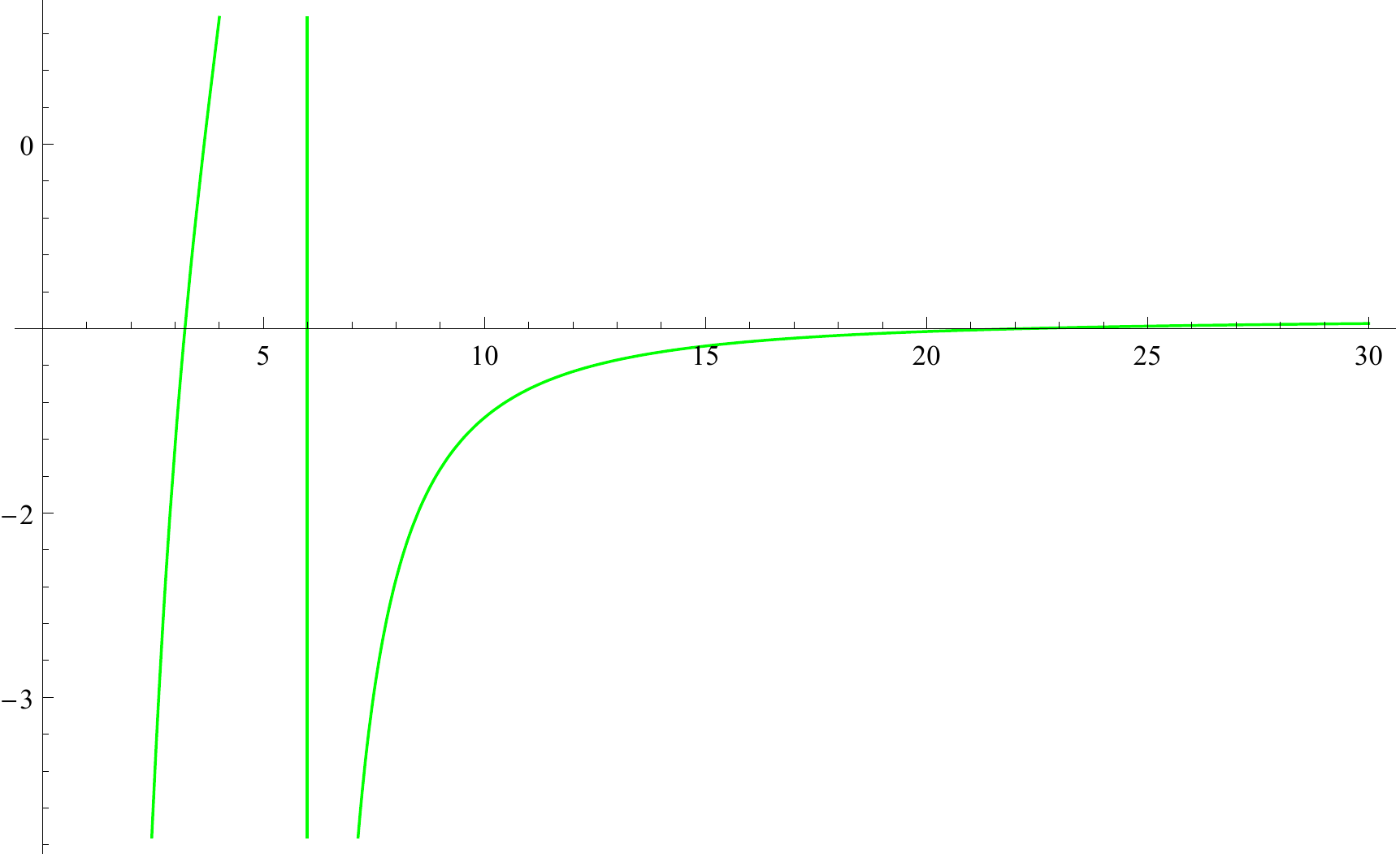}
\end{tabular}
\caption{\scriptsize{Representation of the temporal evolution of $\omega_{DE}(t)$.} }
\label{fig1}
\end{figure}

We now assume an exponential model, as well as \cite{bamba2}, defined by
\begin{eqnarray}
f(\mathcal{T})=\mathcal{T}-\beta \mathcal{T}_s\left(1-\exp\left[\frac{\mathcal{T}}{\mathcal{T}_s}\right]\right)\label{exp}\,.
\end{eqnarray}
We will now test for a solution of the type power law
\begin{eqnarray}
a(t)=t^{\alpha}\,,\,H(t)=\frac{\alpha}{t}\label{plsol}\,.
\end{eqnarray}
{\bf We can show that \eqref{plsol} is a solution of the equations of motion \eqref{H2} and \eqref{Hdot} if the material part is given by the expressions
\begin{eqnarray}
&&\rho_{mt}=\frac{1}{2t^4\mathcal{T}_s\kappa^2}\bigg\{e^{\frac{6\alpha[a_1+\alpha(1-2a_1)]}{\mathcal{T}_st^2}}\left[\mathcal{T}_s^2t^4+6\alpha\mathcal{T}_st^2(a_1+\alpha(3a_1-2))+72a_1\alpha^2(a_1(2\alpha-1)-\alpha)\right]\beta\nonumber\\
&&+\mathcal{T}_st^2\left[-6\alpha^2+6\alpha a_1(2+\alpha)-\beta\mathcal{T}_st^2\right]\bigg\}\\
&&p_{mt}=\frac{1}{2\mathcal{T}_s^2\kappa^2 t^6}\bigg\{e^{\frac{6\alpha[a_1+\alpha(1-2a_1)]}{\mathcal{T}_st^2}}\big[-\mathcal{T}_s^3t^6+2\mathcal{T}_s^2\alpha t^4(-4+3a_1(2-3\alpha)+6\alpha)+288a_1\alpha^2(a_1+\alpha(1-2a_1))^2\nonumber\\
&&-24\mathcal{T}_st^2\alpha(3a_1-\alpha)(-\alpha+a_1(2\alpha-1))\big]\beta+\mathcal{T}_s^2t^4\left[2\alpha(-4-3a_1(\alpha-1)+3\alpha)+\beta\mathcal{T}_s t^2\right]\bigg\}
\end{eqnarray}
} 
The figure \ref{fig1} is the temporal evolution of the state parameter $\omega_{DE}$ of the dark energy. The red curve is obtained with constant given by $\{\alpha=2,\beta=1,\mathcal{T}_s=H_0\Omega_m^{(0)}/\beta,H_0=0.75,\Omega_m^{(0)}=0.23,a_1=1000\}$, where we can see that the fluid is always phantom $\omega_{DE}<-1$. The blue curve is obtained with constant given by $\{\alpha=20,\beta=1,\mathcal{T}_s=H_0\Omega_m^{(0)}/\beta,H_0=0.75,\Omega_m^{(0)}=0.23,a_1=1\}$, where we can see that the fluid is always phantom $\omega_{DE}<-1$, but it fluctuates approximately between the values $-1.05$ and $-1.08$. The most interesting case is the green curve obtained for the constants $\{\alpha=2,\beta=1,\mathcal{T}_s=H_0\Omega_m^{(0)}/\beta,H_0=0.75,\Omega_m^{(0)}=0.23,a_1=0.1\}$. In this case we see that the fluid begins in a rather phantom phase, going through another phase type quintessence, heading toward a behavior of baryonic matter ($\omega>0$) and finally returning the phantom phase. The result is that the current accelerated expansion of the universe and the crossing of the phantom divide from the phantom phase to the non-phantom (quintessence) one can be realized, as well as in \cite{bamba2}.   

\subsection{Thermodynamics for a apparent horizon}\label{subsec6.4}

A further application is for Thermodynamics of the apparent horizon in cosmology FLRW metric. We can follow the formulation given in \cite{ednaldo}. 
\par 
We can establish a similar equation of continuity, deriving over time \eqref{H2} and using \eqref{Hdot}
\begin{eqnarray}
\dot{\rho}_{m}+\dot{\rho}_{DE}+3H\left(\rho_{m}+\rho_{DE}+p_m+p_{DE}\right)=3H^2\frac{d}{dt}\left(\frac{1}{G_{eff}}\right)\label{conserv}\,.
\end{eqnarray}
Whereas the baryonic matter is conserved ($\dot{\rho}_m+3H(\rho_m+p_m)\equiv 0$), we can see that dark energy is not conserved, yielding the interpretation that it is a system out of equilibrium with entropy production (non-equilibrium thermodynamics).  Following exactly the same steps in \cite{ednaldo}, we can establish the first law of thermodynamics
\begin{eqnarray}
T_AdS_A+T_AdS_p=-dE_{MS}+WdV\,,\,T_AdS_p=-\frac{1}{2}\hat{r}_A(1+2\pi  \hat{r}_A T_A )d\left(\frac{1}{G_{eff}}\right)\,.\label{flt}
\end{eqnarray}
at where $T_A$ é the temperature of the apparent horizon, $dS_A$ is the entropy of the apparent horizon, $dS_p$ is the produced entropy, $dE_{MS}$ is the Misner-Sharp energy, $W$ the work and $dV$ the volume element of the apparent horizon. Here it is clearly seen that the first law of thermodynamics is consistent for entropy production associated with an effective Newton constant $G_{eff}$, given in \eqref{H2}, which for the linear case of $f(\mathcal{T})$ the entropy production vanishes and the system back to equilibrium. 
\par 
{\bf If we take the same model of the previous section, i.e. \eqref{exp} and \eqref{plsol}, we can explicitly show the time dependence of effective Newton constant in \eqref{H2}
\begin{eqnarray}
G_{eff}=\left(3a_1-2\right)^{-1}\left\{1+\beta e^{\frac{6\alpha[a_1+\alpha(1-2a_1)]}{\mathcal{T}_st^2}}\right\}^{-1}\label{Geff}
\end{eqnarray}
Here are two important observations. The first is that it becomes explicit dependence of the first law of thermodynamics to the specific choice of the value on $a_1$ in \eqref{Geff}. The second is that by taking the particular value $\beta\equiv 0$ in \eqref{Geff}, clearly we have $G_{eff}=\left(3a_1-2\right)^{-1}$, which again shows the dependence of the theory in relation to the specific value of $a_1$, and from \eqref{Geff}, \eqref{exp} and \eqref{flt} we return to linear theory, where there is no entropy production.
} 
\section{Conclusion}\label{sec7}
We construct a theory that describes the gravitational interaction through effects of torsion of space-time. This theory generalizes the Teleparallel Theory keeping the invariance by both local Lorentz transformations as general coordinates transformations  for a particular case. 
\par
The action of our theory is described by a general algebraic function that depends on a tensorial scalar $\mathcal{T}$ which is classified by a real parameter $a_1$. Our theory falls exactly  in $f(\bar{R})$ Gravity when we take the limit in which $a_1\rightarrow 1$. This is shown from the equations of motion of the two theories.
\par 
We show explicitly through the equations of motion of our theory that it is also equivalent to recent $f(T,B)$ Gravity, when $f(T,B)=f(-T+a_1B)$. 
\par 
We make two small applications of our theory, reconstructing the action for the particular case of de Sitter universe for the flat FLRW metric, with a set of diagonal tetrads, and for obtain a static type-de Sitter solution.  We also analyse the evolution of the state parameter of the dark energy and the first thermodynamics law for the apparent horizon.
\par 
Our theory is a good scenery to an attempt to explain the accelerated expansion of our universe, by modifying the teleparallel usual gravitation, or analogous to Einstein gravity. The real parameter $a_1$ which classifies which theory the GTT describes, it is crucial to any consideration of cosmological phenomena. We also expect new solutions of black holes arise through our theory, in which may also suggest some light on the so-called dark matter explanation on local effects of gravitation.
\par 
{\bf Another perspective is to show the stability of the three solutions discussed here. This should be a topic for future work.}

\vspace{1cm}
{\bf Acknowledgement}: Manuel E. Rodrigues  
thanks UFPA, Edital 04/2014 PROPESP, and CNPq, Edital MCTI/CNPQ/Universal 14/2014,  for partial financial support. 



%

\end{document}